\documentclass[aps,pre, preprint,groupedaddress]{revtex4}
\usepackage{graphicx}
\usepackage{booktabs}
\usepackage{topcapt}
\usepackage{epsfig}
\usepackage{bm}

\begin{document}

\title{Phase diagram and water-like anomalies in core-softened shoulder-dumbbells system}

\author{Cristina Gavazzoni$^1$,Guilherme K. Gonzatti$^2$, Luiz Felipe Pereira$^1$,Luis Henrique Coelho 
Ramos$^1$, Paulo A. Netz$^2$, Marcia C. Barbosa$^1$}

\affiliation{$^1$ Instituto de F\'{\i}sica, Universidade Federal do Rio
Grande do Sul, Porto Alegre, RS, 91501-970, Brazil.\\
$^2$ Instituto de Qu\'imica, Universidade Federal do Rio Grande do Sul,
Porto Alegre, RS, 91501-970, Brazil.}

\date{\today}

\begin{abstract}

Using molecular dynamics we studied the role of the anisotropy on the phase 
boundary and on the anomalous  behavior of 250  dimeric particles interacting by a 
core-softened potential. This study led us to an unexpected result: the introduction 
of a rather small anisotropy, quantified by the distance between the particles 
inside each dimer, leads to  the increase of the size of the regions in the
pressure-temperature phase diagram of anomalies 
when compared to the isotropic monomeric case. However, as the anisotropy 
increases beyond a threshold the anomalous regions shrinks. We found that
this behavior can be understood by decoupling the 
translational and non-translational 
kinetic energy components into different translational and 
non-translational temperatures.

\end{abstract}

\maketitle

\section{Introduction}

There are liquids in nature that exhibit unexpected thermodynamic or
transport properties. Water belongs in this group of anomalous liquids and it is the one with the most 
puzzling behavior. It expands upon cooling at fixed pressure (density anomaly), diffuses faster upon compression 
at fixed temperature (diffusion anomaly) and become less ordered upon increasing density at constant temperature (structural anomaly). 
Despite the complexity of its anomalies, simple molecular water models proved to be able to reproduce many
of water unusual properties \cite{Ne01,Er01,An00}.

The origin of water's anomalies is related to the competition between open low-density and closed 
high-density structures, which depend on the thermodynamic state of the liquid \cite{Kr08}. 
With the purpose of understanding the nature of these anomalies several models were proposed in 
which spherically  symmetric particles interact with each other by an isotropic potential
\cite{Sc00,Fr01,Bu02,Bu03,Sk04,Fr02,Ba04,Ol05,He05,He05a,He70,Ja98,Wi02,Ku04,Xu05,Ne04,Ol06,Ol06a,Ol07,
Ol08,Ol08a,Ol09,Fo08,Al09,Pr04,Gi06} where this competition 
is described by two preferred interparticle distances. 

Among these models, the core-softened shoulder potential proposed by Oliveira \textit{et. al.}\cite{Ol06,Ol06a}
reproduces qualitatively some of the most remarkable water's anomalies. In particular the regions of density anomaly,
diffusion anomaly  and structural anomaly obey the same hierarchy as found in real water \cite{Ne01}.
This potential can represent in 
an effective and orientation-averaged way the interactions between water clusters characterized by the 
presence of those two structures (open and closed) discussed above. 
The open structure is favored by low pressures and the closed structure is 
favored by high pressures, but only becomes accessible at sufficiently high temperatures. 

Even though core-softened potentials have been mainly used for modeling water  \cite{Ol05,Ol06,Ol06a,Ol09,Gi06,Xu06,
Ya05,Ya06,Fr07}, many other materials present the so called  water-like anomaly behavior\cite{Th76, Ts91,Po97,An00,Sh02, 
Sha06,An76,Sa03} and, in principle, could be represented by two length potentials. In this sense, it is  
reasonable  to use core-softened potentials as the building blocks of a broader class of materials
which we can classify as anomalous fluids. The success of these models raise the question of what properties are 
not well represented by spherical symmetric potentials.

In a recently study we compared  the monomeric model proposed by Oliveira \textit{et. al.}\cite{Ol06,Ol06a} with
a model of dimeric molecules linked as rigid dumbbells, interacting with the same potential. The introduction
of anisotropy leads to a much larger region of solid phase in the phase diagram and to the appearance of a 
liquid crystal phase\cite{Ol10}. Later we showed that the size of anomalous regions in the pressure temperature 
phase diagram is dependent on the value of interparticle separation $\lambda^*$ and that the effect of the 
introduction of a rather small anisotropy due to the dimeric nature of the particle leads to the increase of 
the size of the regions of anomalies. Nevertheless, the increase of $\lambda^*$ shrinks those regions \cite{Ne11}. 

A possible explanation for this unexpected behavior is that, in conditions of low temperature and pressure,
there is a decoupling of translational and rotational degrees of freedom when the interparticle separation is small.
Experimental works showed that when approaching the glass transition the translational and rotational dynamics 
are not equally affected \cite{Ha97,Ci96}. This behavior was also observed in computer simulations \cite{Se12}.

In order to test this hypothesis here we study a dimeric model where each particle in the dimer interacts with 
the particles in the other dimers through a core-softened potential. We define  translational and
non-translational temperatures as tools to evaluate the role of translational and non-translational degrees
of freedom. Different values of the interparticle separation $\lambda^*$ were used and compared.

The remaining of this manuscript goes as follows. In Sec. II the model is introduced and the simulation 
details are presented. In Sec. III the results are shown and conclusions are exposed in Sec. IV.

\section{The model}

\begin{figure}[h!tb]
\includegraphics[clip=true,scale=0.3]{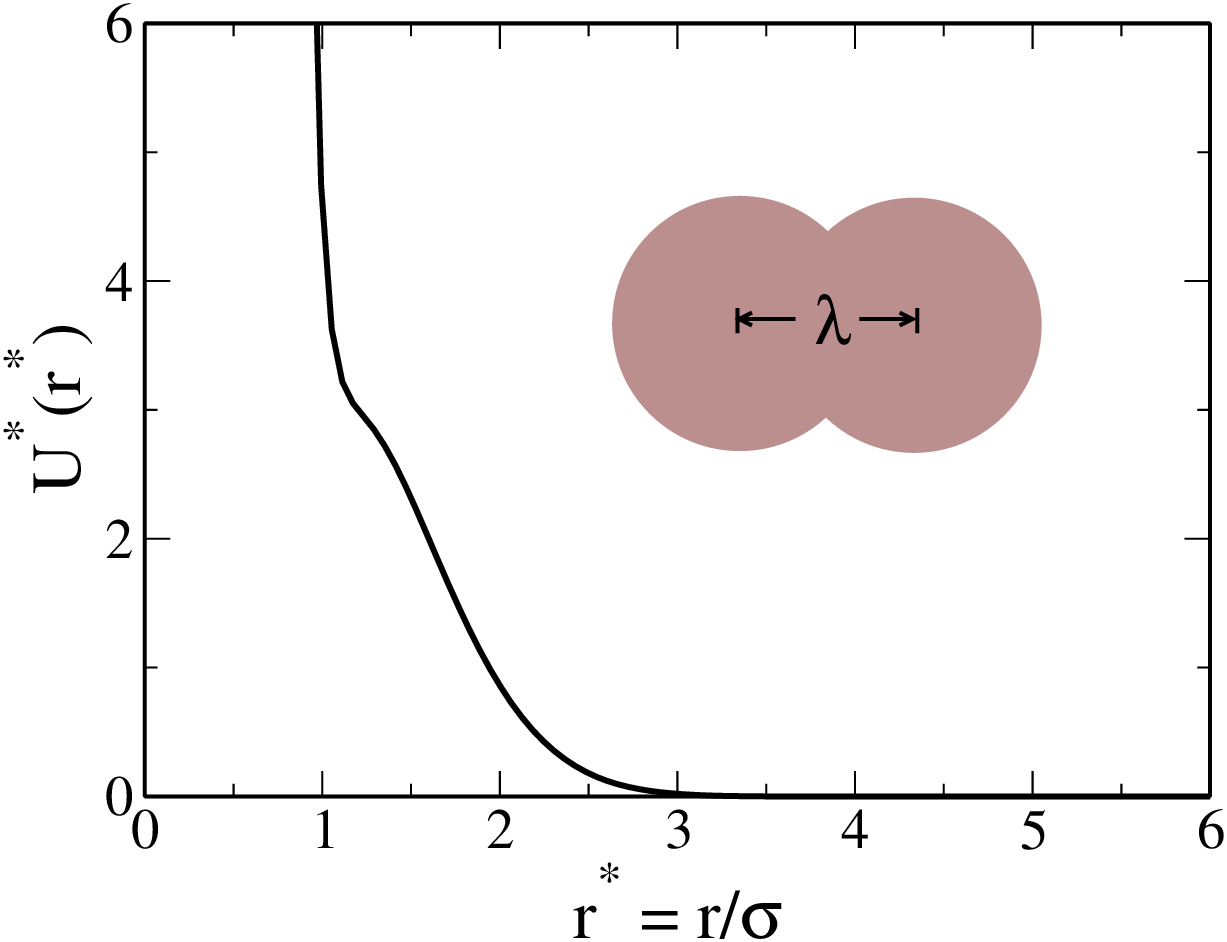}
\caption{Effective potential versus distance in reduced units.}
\label{potencial}
\end{figure}

Our model consists of N spherical particles of diameter $\sigma$, linked rigidly in 
pairs with the distance $\lambda^*$ between their centers of mass forming dimers as shown
in figure \ref{potencial}. In this way, $\lambda^*$ is related to the anisotropy of the dimers. 
Each particle of the dimer interacts with all particles belonging to other dimers
with the intermolecular continuous shoulder potential \cite{Ol06,Ol10} given by

\begin{equation}
U^{*}(r)= \frac{U(r)}{\epsilon}=4\left[\left(\frac{\sigma}{r}\right)^{12}-
\left(\frac{\sigma}{r}\right)^{6}\right]+
a\exp\left[-\frac{1}{c^{2}}\left(\frac{r-r_{0}}{\sigma}
\right)^{2}\right]\; .
\label{pot}
\end{equation}  
\\

Depending on the choice of the values of $a$, $r_0$ and $c$ a whole family of
potentials can be built, which shapes ranging from a deep double wells\cite{Ne04,Th76} to a repulsive 
shoulder\cite{Ja98}. In our simulations we used $a=5$, $r_0/\sigma=0.7$, $c=1$ and $\lambda/\sigma = 0.10, 0.20, 0.50$ and $0.70$.

We performed molecular dynamics simulations in the canonical ensemble using $N=500$
particles (250 dimers) in a cubic box of volume $v$ with periodic boundary conditions 
in three directions, interacting with the intermolecular potential described above. 
The cutoff radius was set to $5.5\sigma$ length units.
Pressure, temperature, density and diffusion are calculated in dimensionless units

$$
T^*\equiv \frac{k_BT}{\epsilon}
$$

$$
\rho^*\equiv \rho\sigma^3
$$

$$
P^*\equiv \frac{P\sigma^3}{\epsilon}
$$

$$
D^*\equiv \frac{D(m/\epsilon)^{1/2}}{\sigma}
$$

$$
t^*\equiv t\left(\frac{\varepsilon}{m\sigma^2}\right)^{1/2}
$$

$$
\lambda^*\equiv \frac{\lambda}{\sigma}
$$
\\

Thermodynamic and dynamic properties were calculated over 700 000 steps after previous 300 000
equilibration steps. The time step was 0.001 in reduced units, the time constant of the Berendsen thermostat 
\cite{Be84} was 0.1 in reduced units. The internal bonds between the particles in each dimer 
remain fixed using the algorithm SHAKE \cite{Ry77} with a tolerance of $10^{-12}$ and maximum 
of 100 interactions for each bond.

The structure of the system was characterized using the intermolecular radial distribution function, 
$g(r)$ (RDF), which does not take into account the correlation between particles belonging to the same dimers.
The diffusion coefficient was calculated using the slope of the least square fit to the linear part
of the mean square displacement, $<r^2(t)>$ (MSD), averaged over different time origins. 
Both the $g(r)$ and $<r^2(t)>$ where computed relative to the center of mass of a dimer.
The phase boundary between the solid and fluid phases was mapped by the analysis of the change of 
the pattern of mean square displacement and radial distribution function \cite{Ne11}.

In a previous work \cite{Ol10} 
we showed that, depending on the chosen temperature and density, 
the system could be in a fluid phase metastable with respect to the solid phase. In order to locate 
the phase boundary two sets of simulations were carried out, one
beginning with molecules in  a ordered crystal structure and other beginning with molecules 
in a random, liquid, starting structure obtained from previous equilibrium simulations. 

We define the structural anomaly region as the region where
the translational order parameter $t$ given by

\begin{equation}
t=\int_{0}^{\xi_c}|g(\xi)-1|d\xi
\end{equation}
\\

decreases upon increasing density. Here $\xi\equiv r\rho^{1/3}$ is the distance $r$ scaled in units of mean
interparticle separation, computed by the center of mass of the dimers, $\xi_c$ is the 
cutoff distance set to half of simulation box times $\rho^{1/3}$ \cite{Ol06a} and $g(\xi)$ is the radial
 distribution function as a function of the scaled distance $\xi$ from a reference particle. For an ideal 
 gas $g = 1$ and $t = 0$. In the crystal phase $g \neq 1$ over long distances and $t$ is large.

The dynamic of the system was analyzed through the mean square displacement (which allows to calculate the diffusion coefficient) and  
the velocity and orientational autocorrelation function.
The velocity autocorrelation function is calculated  taking the average for all dimers and initial times through
the expression

\begin{equation}
v_{acf}(t)=  \langle \mathbf{v}_i(t_0)\cdot\mathbf{v}_i(t_0 + t) \rangle
\end{equation}
\\

where $\vec{v}_i(t_0)$ is the initial velocity vector for the dimer $i$ and $\vec{v}_i(t_0 + t)$ is the velocity vector
for the same dimer at a later time $t_0+t$. The orientational autocorrelation function is defined by

\begin{equation}
O^l_{acf}(t)=  \langle P_l(cos\theta_i) \rangle 
\end{equation} 
\\

where $P_l$ is the Legendre polynomials of order $l$ and $\theta_i$ is the reorientation angle of the intra-dimer vector during the 
time interval $\Delta t$.

\section{Results}

\subsection{P-T phase diagram for different values of $\lambda^*$}

In previous work \cite{Ol10}, we studied a system of 250 dimer, interacting through the interparticle potential (\ref{pot})
with $\lambda^* = 0.20$ in a broad range of densities and temperatures. The pressure, radial distribution function, mean 
square displacement and diffusion coefficient were calculated for all state points, yielding a complete description 
of the regions of structural, density and diffusion anomalies.

The comparison between the monomeric case and the dimeric case with $\lambda^*=0.20$ showed
in the pressure vs temperature phase diagram
that the anisotropy due the dumbbells leads to 
larger range of pressures and temperatures occupied by anomalous regions and by the solid phase \cite{Ol10}. 

\begin{figure}[h!tb]
\includegraphics[clip=true,scale=0.3]{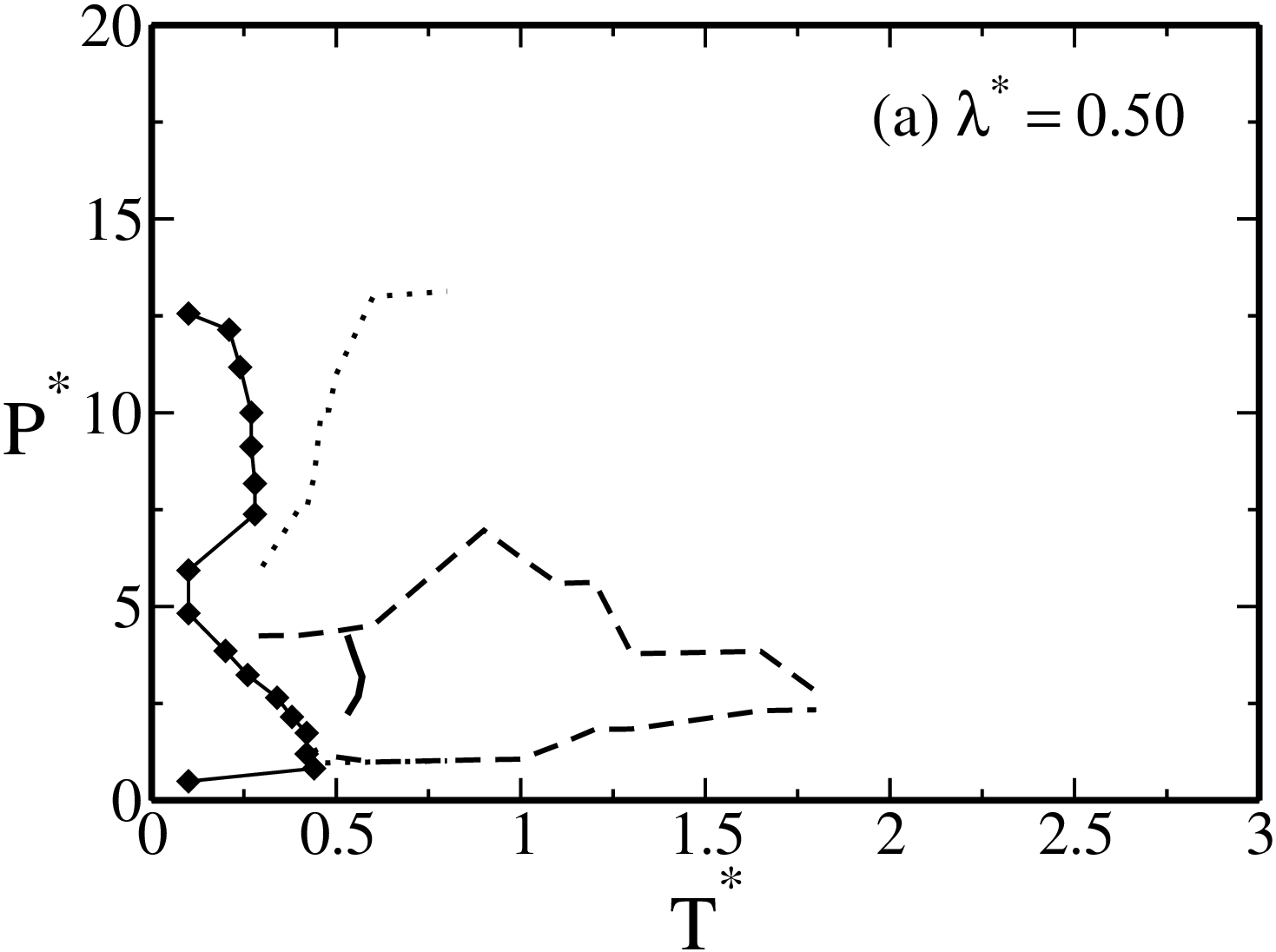}
\includegraphics[clip=true,scale=0.3]{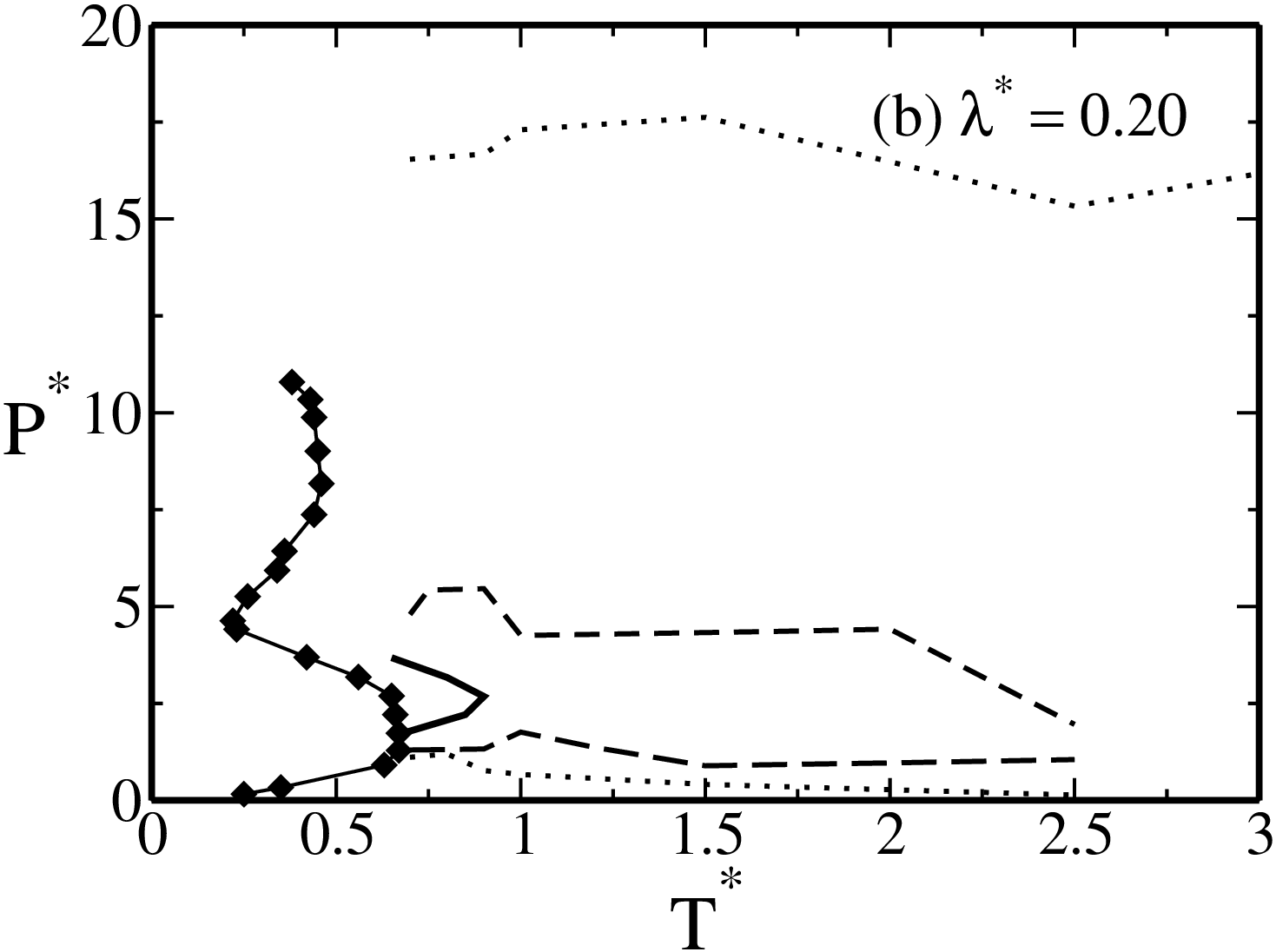}
\caption{Pressure temperature phase diagram for (a)$\lambda^*=0.50$ and (b)$\lambda^*=0.20$ interacting through the potential illustrated in
Fig. \ref{potencial}. The bold line with diamonds represent the boundary between the fluid and solid phases 
and the bold, dashed and dotted lines represent the TMD, the region of diffusion anomaly and the region of structural 
anomaly respectively.}
\label{pt}
\end{figure}

This result raises questions about the influence of anisotropy on the phase and anomalous behavior.
In principle one could expect that by increasing the anisotropy by enlarging $\lambda^*$ the
solid-fluid phase boundary, the TMD line and the diffusion anomalous region would move to higher temperatures.
In order to check this assumption we carried out a detailed set of simulations of rigid dumbbells 
interacting with the potential described above with $\lambda^*=0.50$ \cite{Ne11}.  Figure \ref{pt} shows 
the complete pressure-temperature phase diagram for this system and for the system with  
$\lambda^* = 0.20$. This figure shows that the boundaries of the 
anomalous regions depend clearly on $\lambda^*$. The solid-fluid phase boundary
and the TMD are shifted towards lower temperatures and slightly higher pressures with the increase 
of $\lambda^*$. The high-temperature limit of the boundary of the region of diffusion anomalies also
becomes shifted towards lower temperatures, whereas the lower-temperature part of this region 
becomes broader. The structurally anomalous region shrinks with increasing $\lambda^*$.

\begin{figure}[h!tb]
\includegraphics[clip=true,scale=0.4]{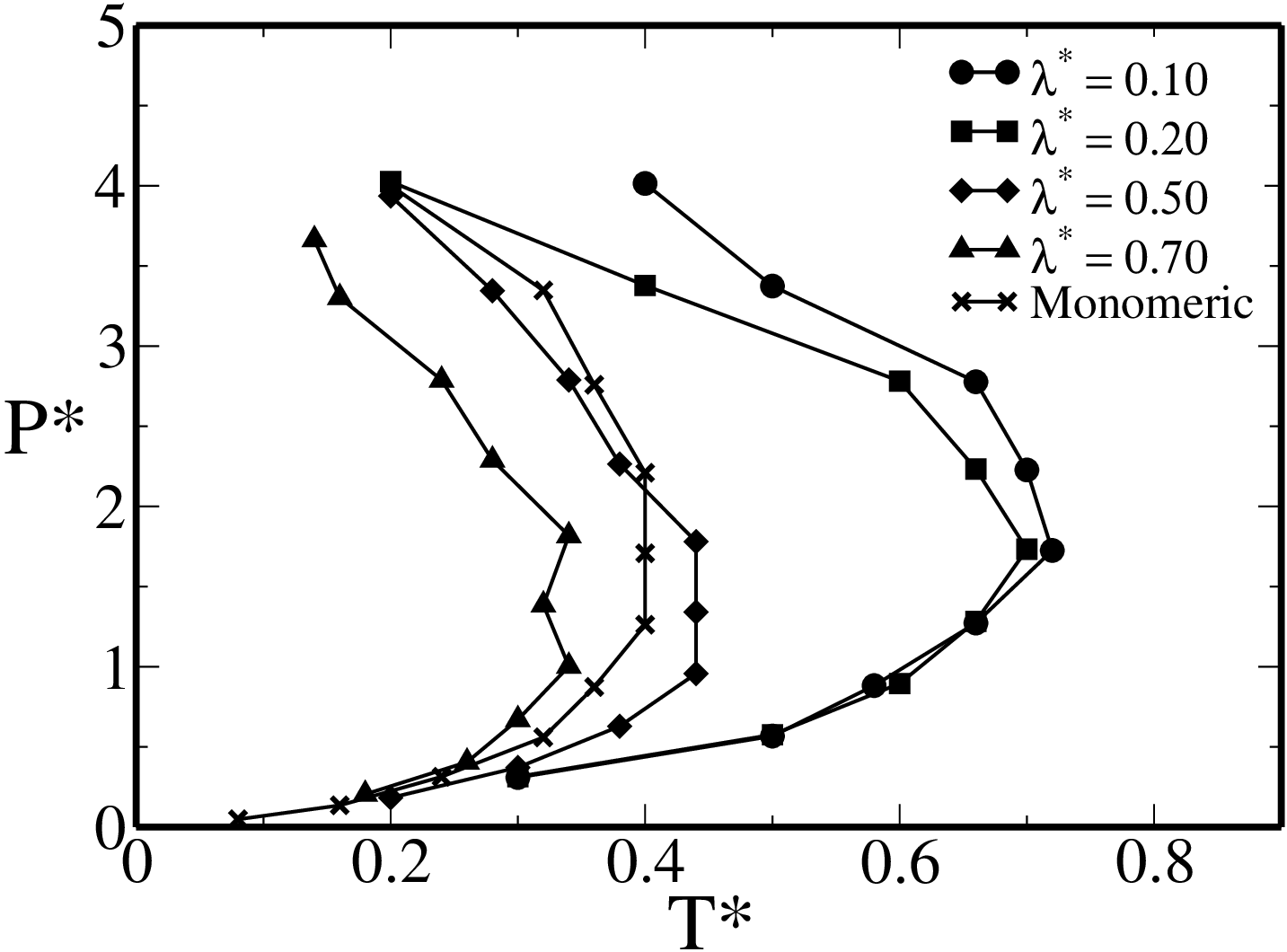}
\caption{Solid-fluid phase boundary for different values of interparticle separation $\lambda^*$ and for monomeric 
case\cite{Ol06, Ol06a}.}
\label{fases}
\end{figure}

In order to confirm the influence of $\lambda^*$ on the phase diagram and regions of anomalous behavior,
several simulations restricted to a small region in the phase diagram were carried out. These simulation explored the 
densities between $\rho^*=0.10$ and $\rho^*=0.30$ and the temperatures betwenn $T^*=0.10$ and $T^*=0.80$ for 
$\lambda^* = 0.10, 0.20, 0.50$ and $0.70$. 
Figure \ref{fases} the solid-fluid phase boundary in the pressure temperature phase diagram for
$\lambda^* = 0.10, 0.20, 0.50$ and $0.70$. The figure shows that the increase 
of the interparticle separation  shifts the  solid-fluid phase boundary to lower temperatures.

Comparing the results of the simulations corresponding to the several values of $\lambda^*$ with the 
results of the monomeric shoulder-potential simulations\cite{Ol06,Ol06a}, a non-monotonic 
behavior is observed instead of having the monomeric case as the limit of $\lambda^*\rightarrow 0$. 
In order to check why as $\lambda^*\rightarrow 0$, the system does not converge to the 
monotonic case, the orientational degrees of freedom are analyzed.

\begin{figure}[h!tb]
\begin{center}
\includegraphics[clip=true,scale=0.3]{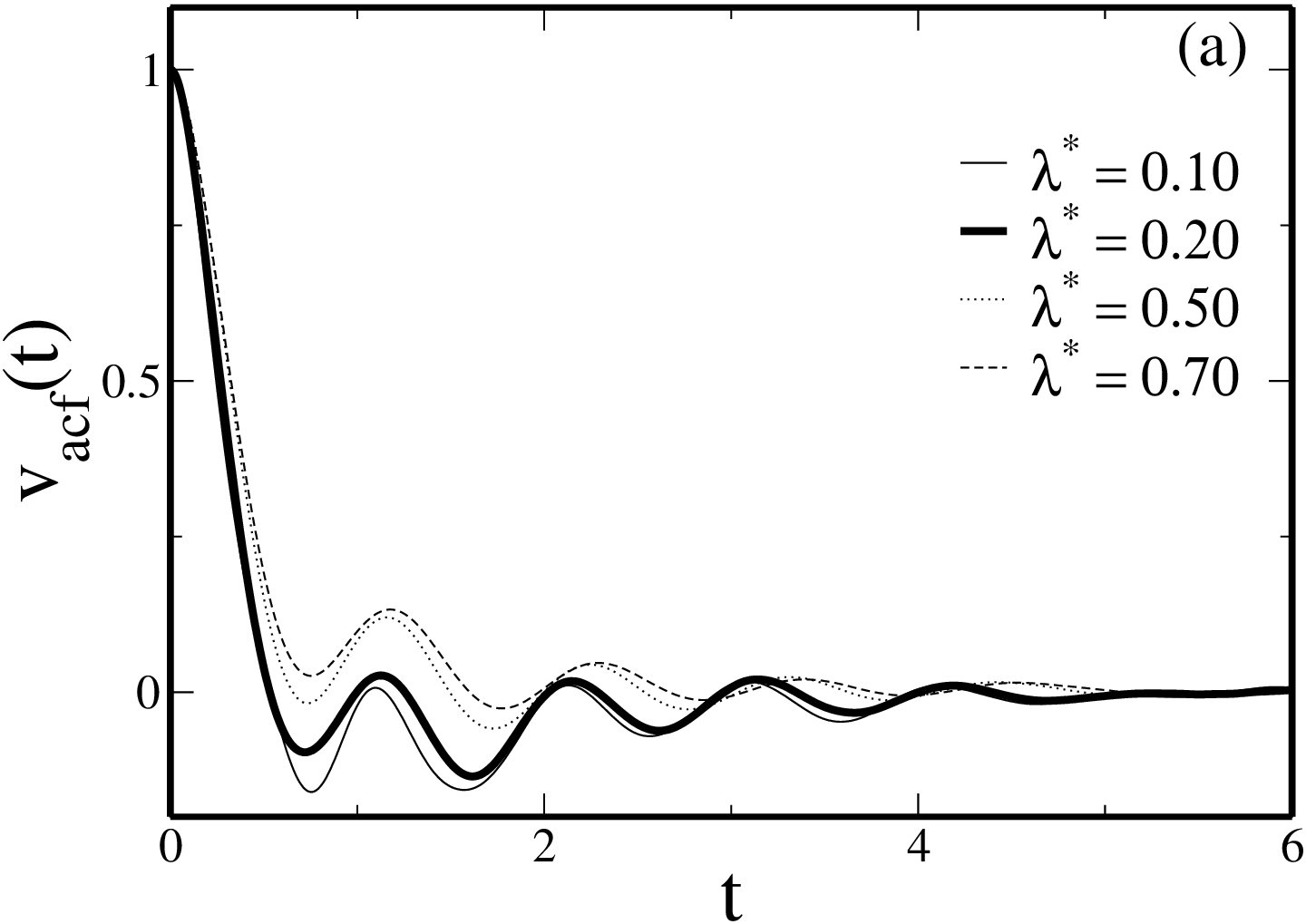}
\includegraphics[clip=true,scale=0.3]{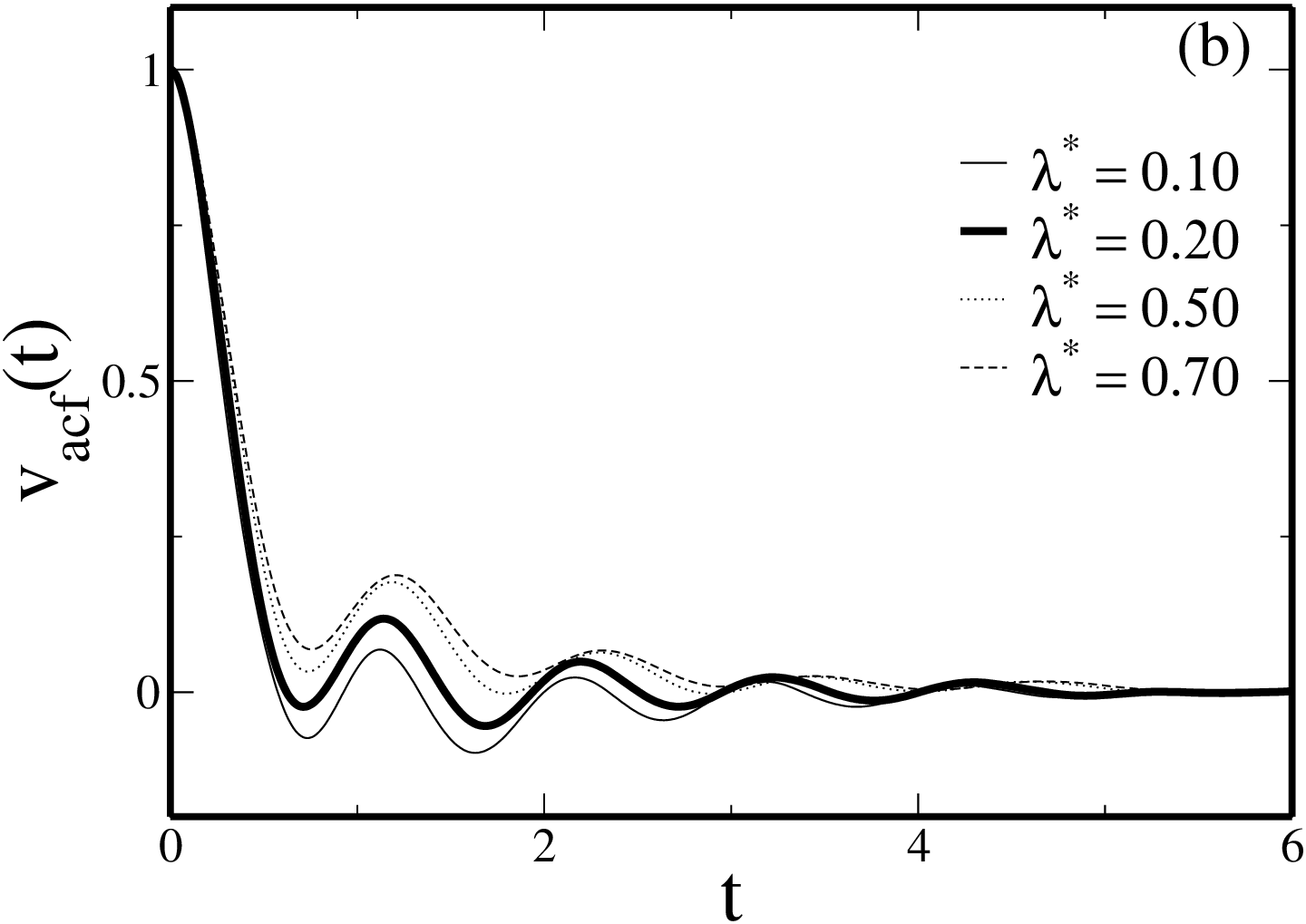}
\includegraphics[clip=true,scale=0.3]{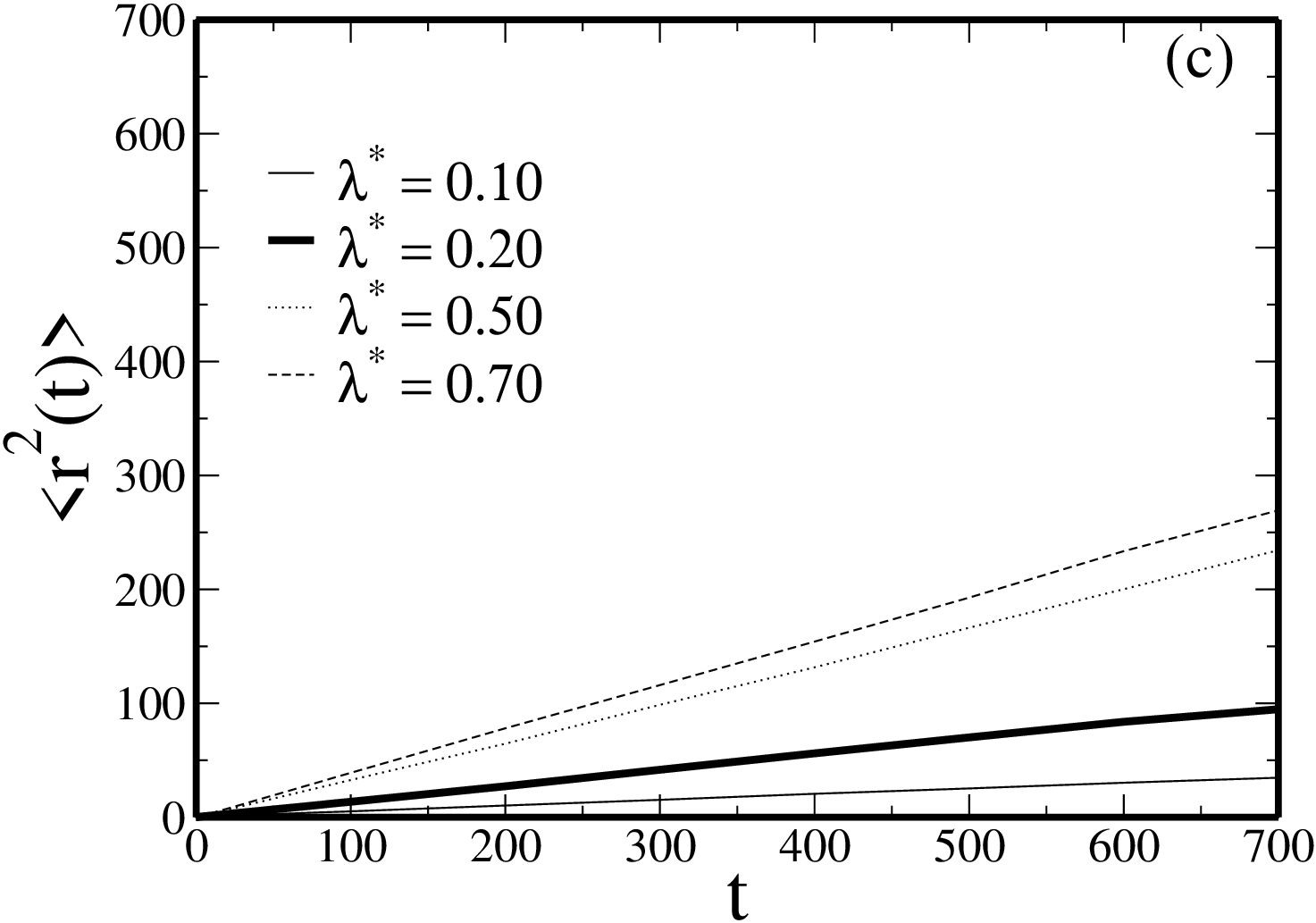}
\includegraphics[clip=true,scale=0.3]{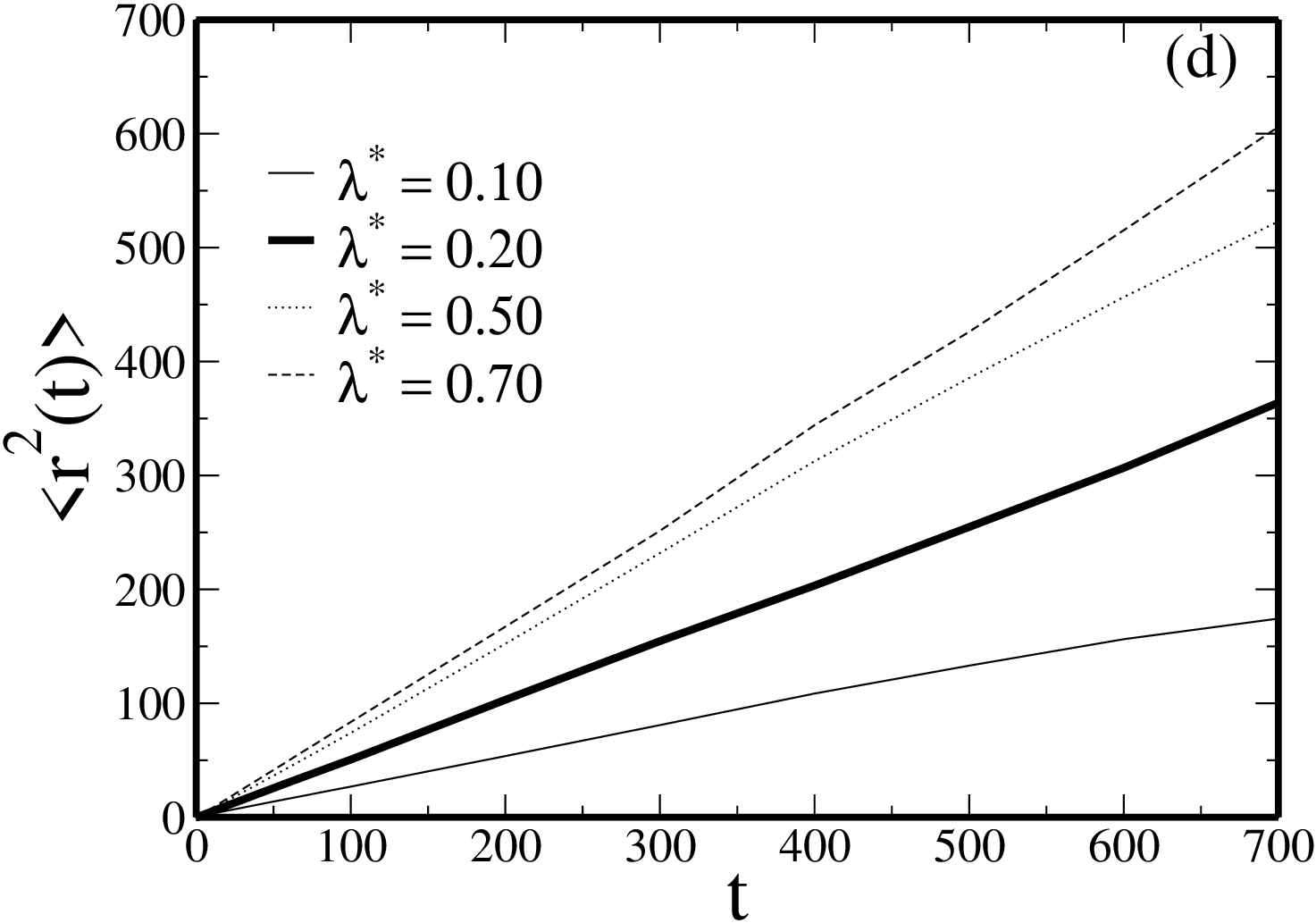}
\includegraphics[clip=true,scale=0.3]{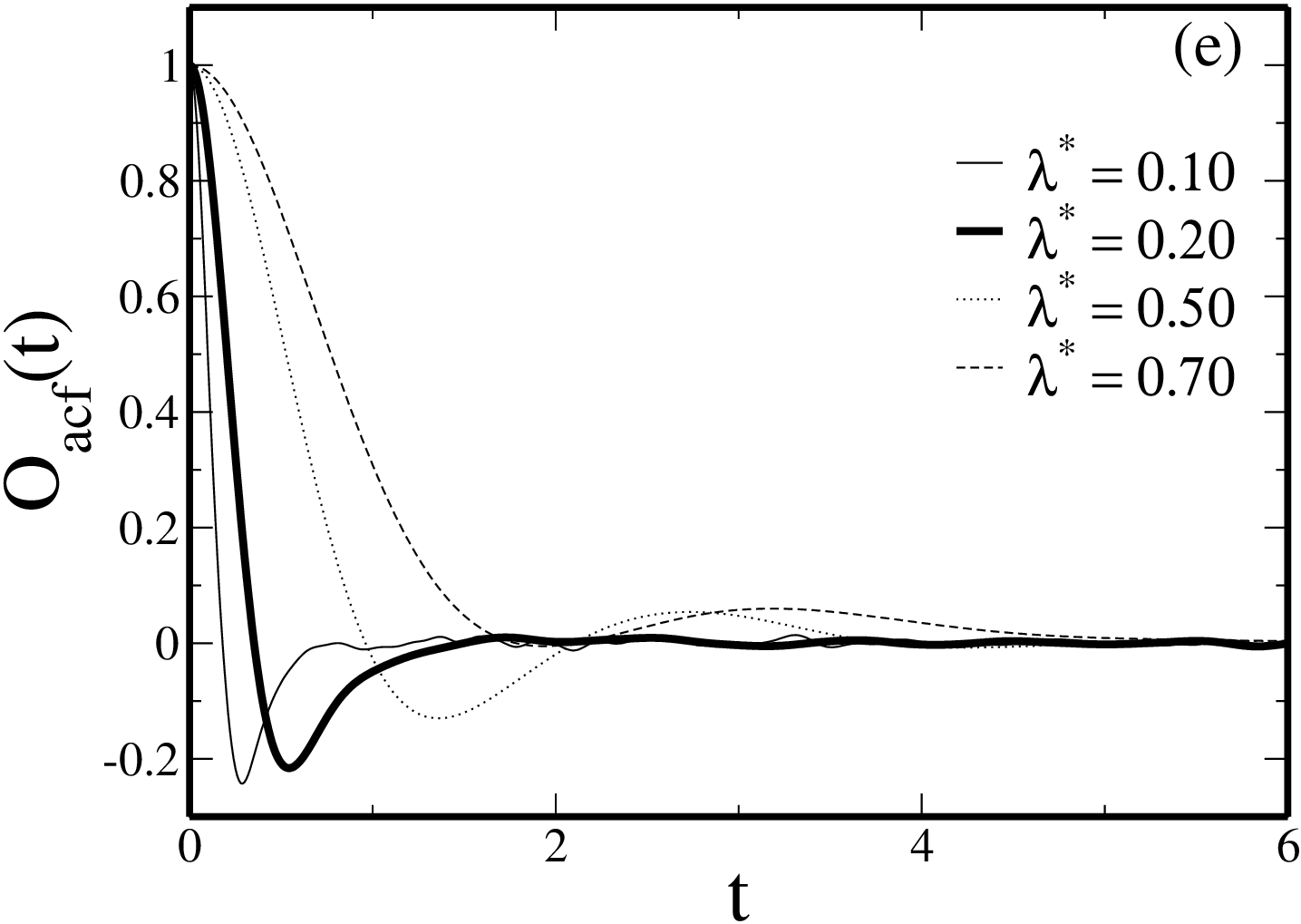}
\includegraphics[clip=true,scale=0.3]{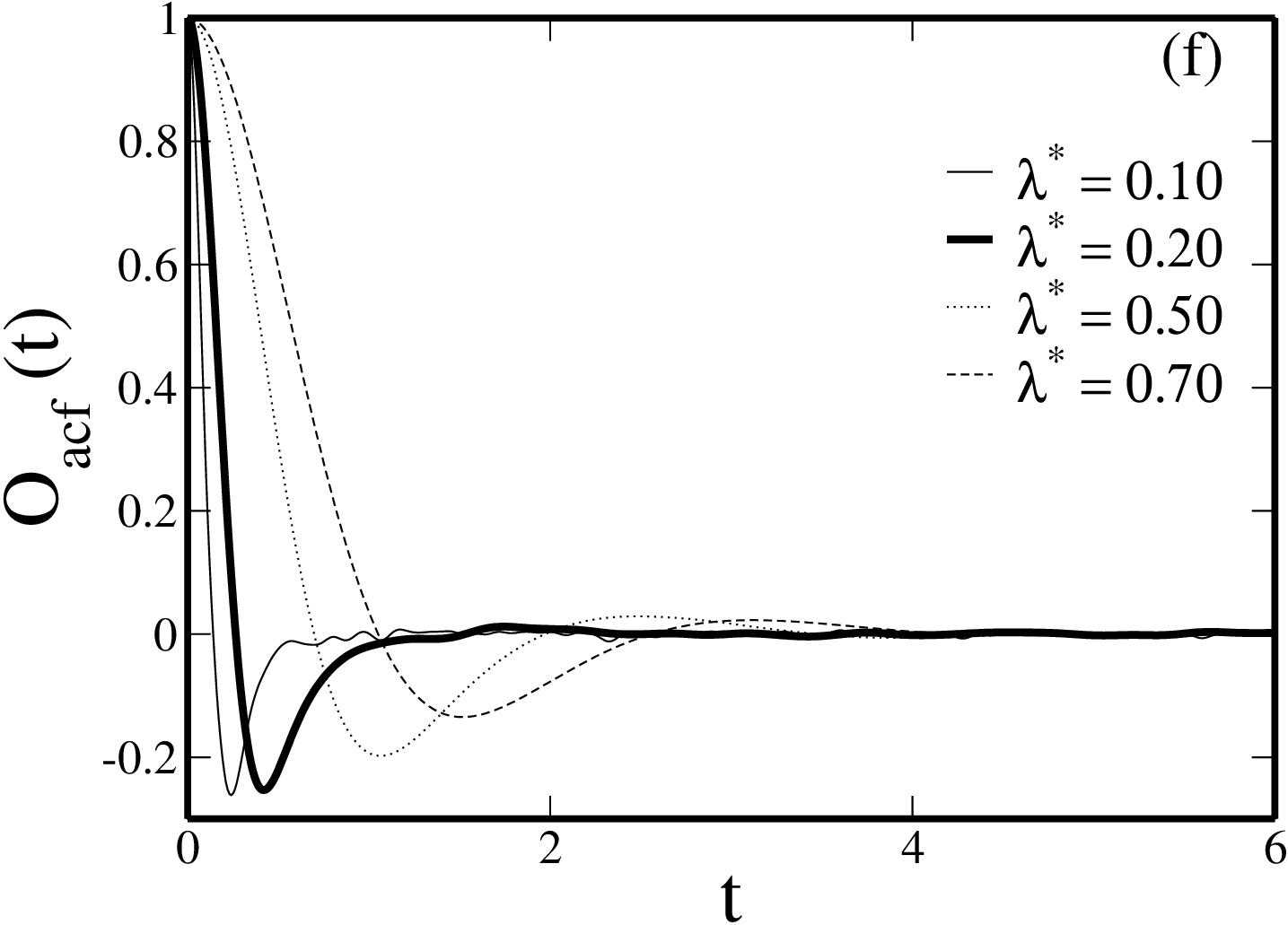}
\caption{Velocity autocorrelation function, $v_{acf}(t)$ for (a)$T^* = 0,30$ and (b)$T^* = 0,50$, mean square displacement, 
$<r^2(t)>$ for (c)$T^* = 0,30$ and (d)$T^* = 0,50$, and orientational autocorrelation function, $O_{acf}(t)$ for 
(e)$T^* = 0,30$ and (f)$T^* = 0,50$. All the graphs are for $\rho^* = 0.30$.}
\label{acf}
\end{center}
\end{figure}

Autocorrelation functions are very powerful tools to describe dynamics therefore the analysis of these curves
can give us an idea why the solid-fluid phase boundary is contracting with the increase of $\lambda^*$. Fig.
\ref{acf} shows velocity autocorrelation functions, $v_{acf}(t)$, orientational autocorrelation function,
$O_{acf}(t)$, and mean square displacement, $<r^2(t)>$, for simulations with different values of $\lambda^*$ subject to the same
conditions of density and temperature.

The velocity autocorrelation function $v_{acf}$ versus time, illustrated in Fig. \ref{acf}, shows that for all 
the values of $\lambda^*$  the curves present a fast decay, however, only the curves for small
interparticle separation $\lambda^*$ cross the zero axis at short time, i.e, only the velocity vectors 
for small $\lambda^*$'s change signal. This result indicate that the less anisotropic 
particles (small $\lambda^*$) "feel"the cage of neighbors molecules more strongly than the more anisotropic 
particles. With the increase of the temperature this effect is weakened, consequently  the behavior of $v_{acf}$
for $\lambda^* = 0.10$ and $T^* = 0.50$ is similar to the behavior of $\lambda^* = 0.20$ and $T^* = 0.30$. Therefore 
the increase of the temperature compensates the decrease of the $\lambda^*$ value.

The slope of mean square displacement, illustrated in Fig. \ref{acf}, is related with the translational diffusion. 
For density $\rho^* = 0.30$ and $T^*= 0.30$, the larger the interparticle separation, $\lambda^*$, the larger 
the translational diffusion of the system would be. With the increase of  the temperature all systems become more diffusive, 
however, the largest change occurs for the small values of $\lambda^*$ . At a fixed temperature thehe relative difference 
between the curves given by:
\\ 
$$
\frac{<r^2(t=700, \lambda^* = 0.70, T^* = 0.30)>}{<r^2(t=700, \lambda^* = 0.10, T^* = 0.30)>}  = 7.77
$$

$$
\frac{<r^2(t=700, \lambda^* = 0.70, T^*=0.50)>}{<r^2(t=700, \lambda^* = 0.10,T^*=0.50)>}  = 3.47
$$ 
\\

The orientational autocorrelation function, $O_{act}$, is also shown in the Fig. \ref{acf}. The less anisotropic systems 
for $\rho^* = 0.30$ and $T^* = 0.30$ have a $O_{act}$ that decays quickly to zero.This result indicates that, 
for these systems, the time needed to perform a $90^o$ rotation is very small, i.e., systems with small 
interparticle separation rotate much easier then systems with larger $\lambda^*$ values. For all the interparticle 
separations, $\lambda^*$, the orientational mobility is  facilitated by the increase of the temperature.
It is noteworthy that the differences in $O_{act}$ (Fig. \ref{acf}(e) e Fig. \ref{acf}(f)) are much
stranger that the difference in $v_{acf}$ (Fig. \ref{acf}(a) e Fig. \ref{acf}(b)).

The analysis of the behavior of $<r^2(t)>$, $v_{acf}$ and $O_{act}$ for several values of $\lambda^*$, $\rho^*$ and $T^*$ 
leads to the conclusion that the more anisotropic systems diffuse faster than the less anisotropic
particles but rotate slower than systems with 
low anisotropy (small $\lambda^*$). Based on these observation we raised the hypothesis that the shrinking of 
the region of solid phase is related to a decoupling between translational and non-translational motions.

In order to test our hypothesis we defined a 
translational temperature and a non-translational temperature as tools to evaluate the role of translational 
and non-translational degrees of freedom.

\subsection{Translational and non-translational temperatures}

In our dimeric system simulations, as well as in our monomeric simulations \cite{Ol06,Ol06a}, 
Berendsen thermostat is used to maintain the temperature fixed. In this thermostat the temperature of the
system is defined as the kinetic energy of the system divided by the number of degrees of freedom. When we are dealing 
with monomeric systems the temperature has only the contribution of the translational kinetic energy but 
when dealing with the dimeric systems, the temperature has the translational plus the 
rotational (and eventually vibrational) kinetic energy contributions. In other words, due to its rotational or vibrational
kinetic energy, a dimeric particle rotating or vibrating around a single point has a non zero temperature, even 
though its center of mass remains still. 
Since the dimeric system is composed by hard-linked particles the internal vibration is excluded, 
remaining only the rotational and hindered rotational (libration) motions.

In order to test the effect in the pressure versus temperature phase diagram
of the decoupling of the temperature we calculated separately the contributions of translational movement, which 
we associate with a translational temperature, and non-translational movement, associated with 
a non-translational temperature. Each degree of freedom contribute with $k_BT/2$ to the kinetic energy. 

The contributions of translational movement were obtained calculating the kinetic energy of the center
of mass of the dimeric particles, $K_{trans}$. Thus if the dimeric particle is rotating around a single point the
translational contribution will be zero. Since there are $N/2$ dimers, the translational temperature is defined by

\begin{equation}
K_{trans} = 3\frac{N}{2}\frac{k_BT_{trans}}{2} 
\end{equation}  
\\

Non-translational temperature was also defined using the equipartition theorem. The internal energy of our full
system is given by 

\begin{equation}
K= f\frac{k_BT}{2} = 5\frac{N}{2}\frac{k_BT}{2}
\end{equation}
\\
where $f$ is the number of degree of freedom of the system. $f$ is
equal to the number of dimers $N/2$ times the number of degrees os freedom of
each dimer. One degree of freedom 
of the dimer is lost due the internal bonds between the particles that belong to the dimer. The 
non-translational  contributions of internal energy is

\begin{equation}
K_{ntrans} = \frac{N}{2}k_BT_{ntrans}
\end{equation}
\\
where  $T_{ntrans}$ is the non-translational temperature.
Using this equation and considering $K = K_{trans}+K_{ntrans}$ we can write an expression for the non-translational temperature

\begin{equation}
T^*_{ntrans}= \frac{5}{2}T^* - \frac{3}{2}T^*_{trans}
\end{equation}
\\

\begin{figure}[h!tb]
\begin{center}
\includegraphics[clip=true,scale=0.3]{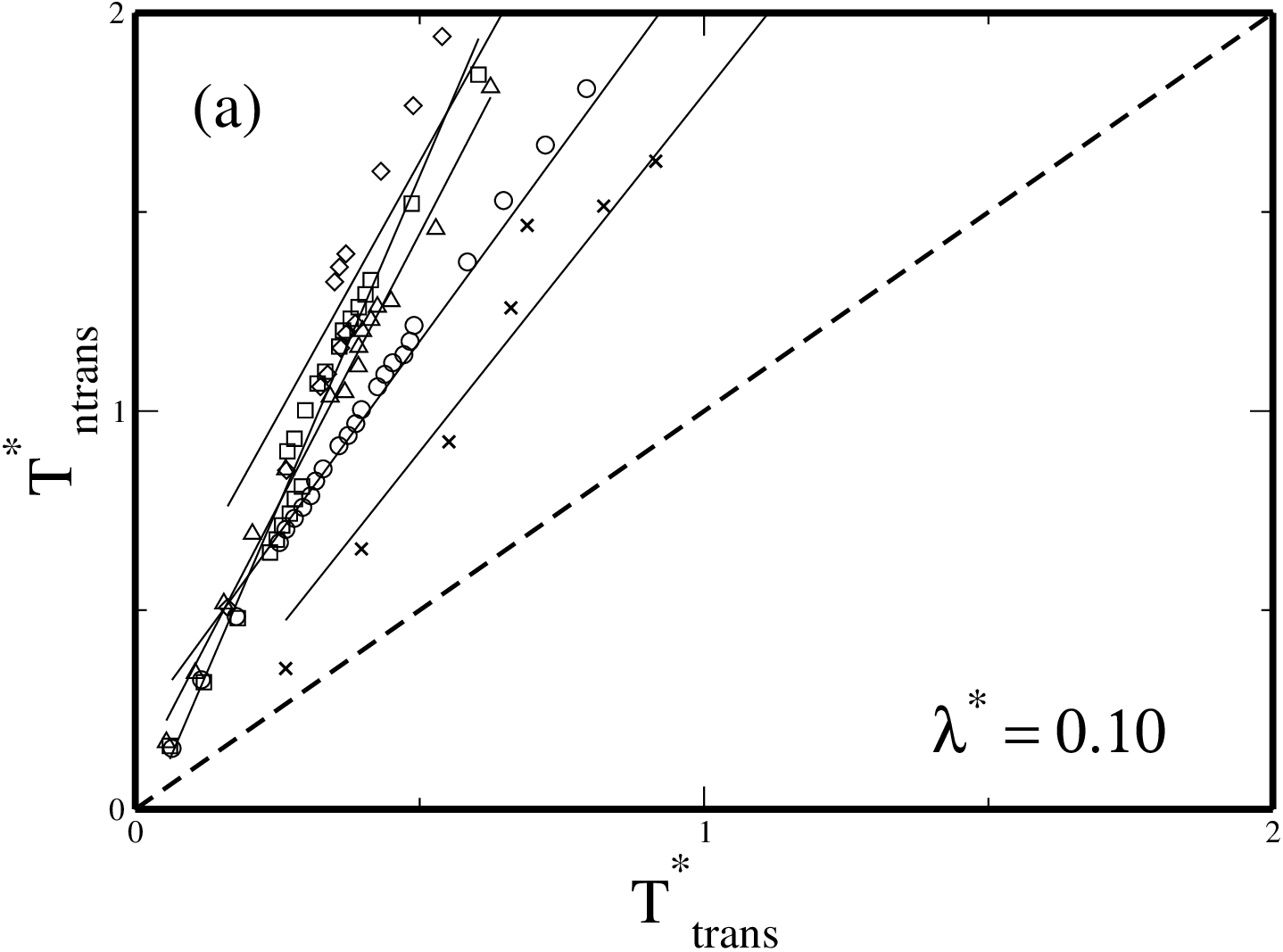}
\includegraphics[clip=true,scale=0.3]{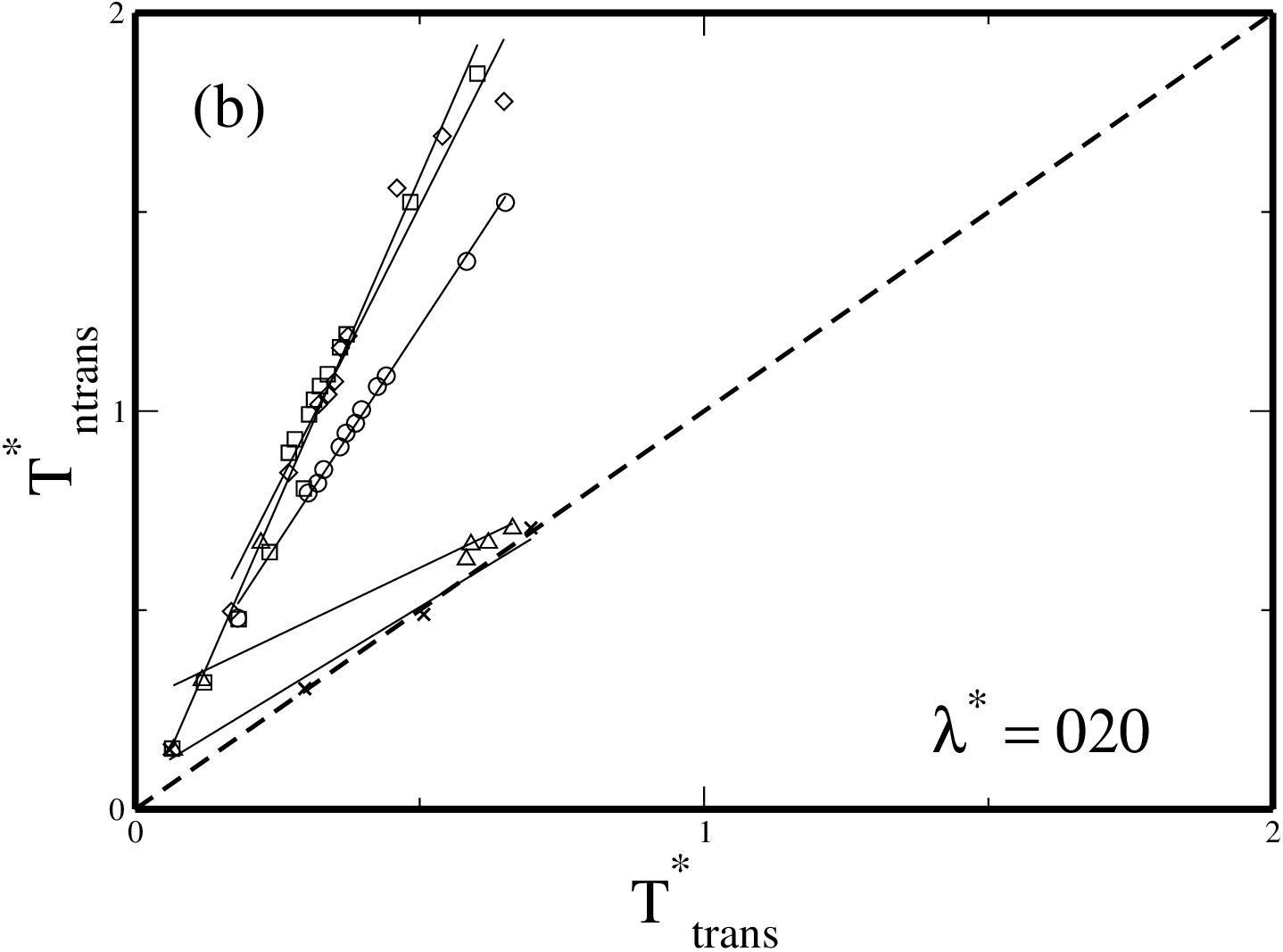}
\includegraphics[clip=true,scale=0.3]{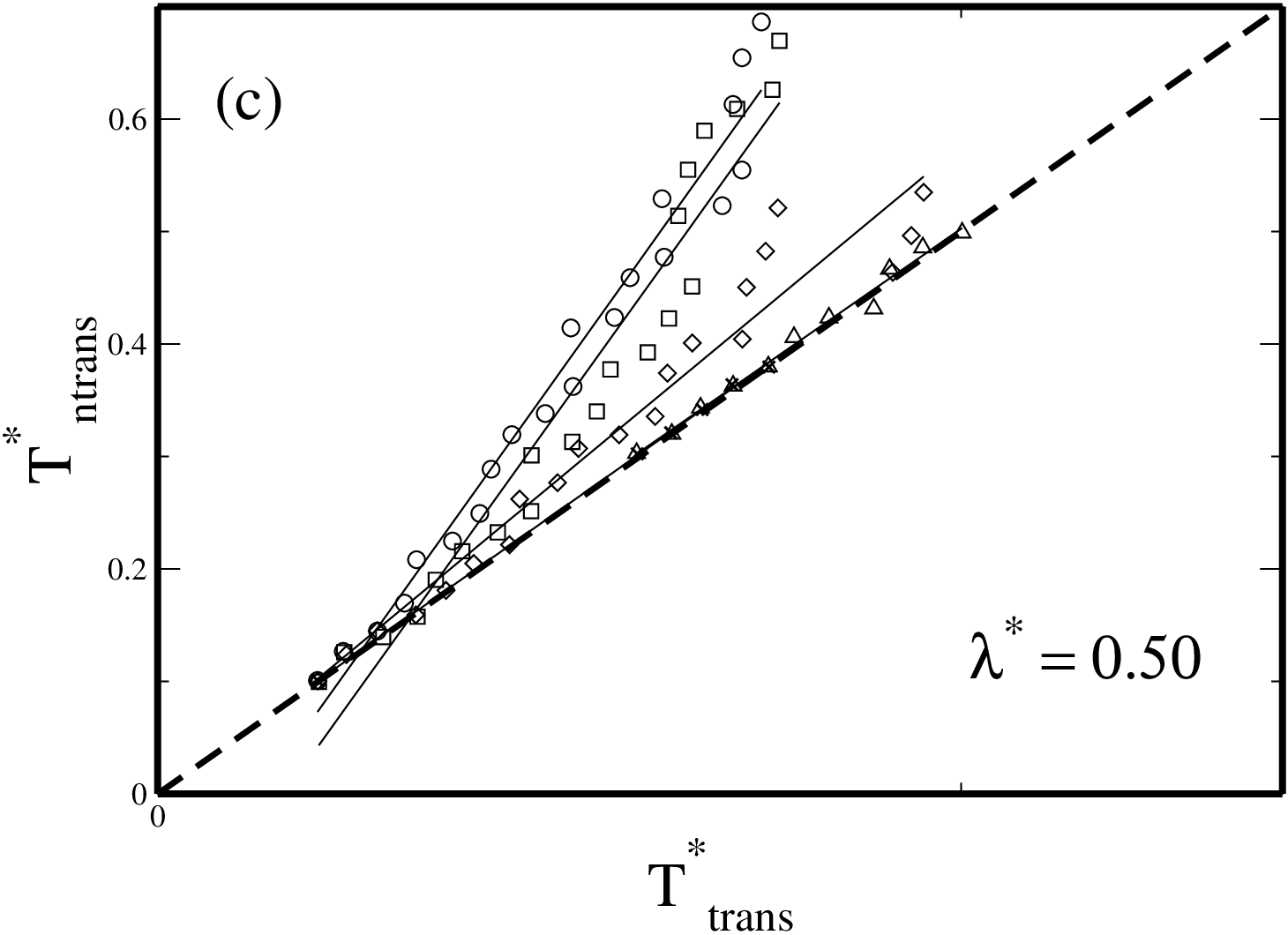}
\includegraphics[clip=true,scale=0.3]{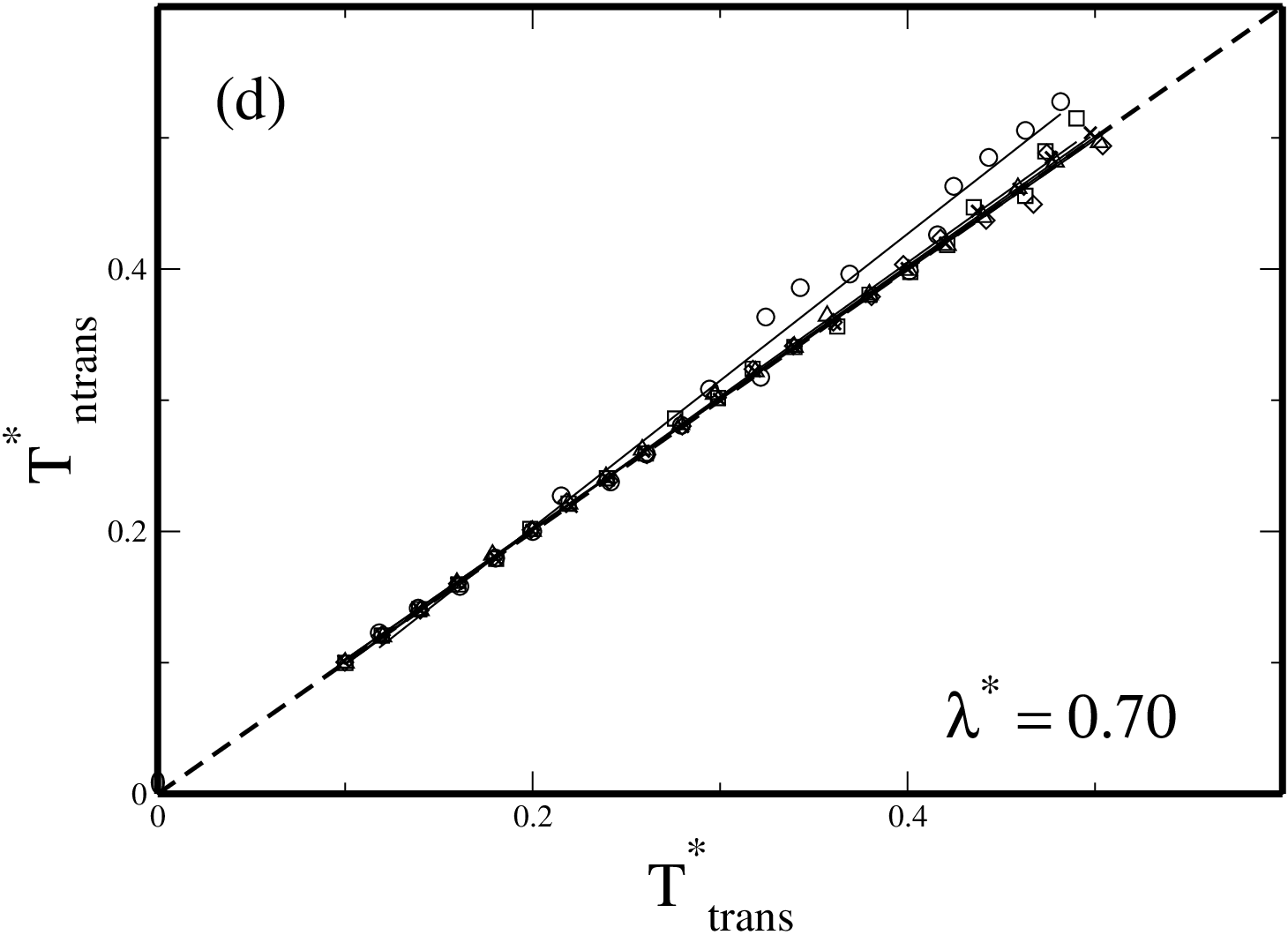}
\caption{Translational temperature vs non-translational temperature for  $(a)\lambda^*=0.10, (b)\lambda^*=0.20, (c)\lambda^*=0.50$ and
$(d)\lambda^*=0.70$. In each case results with  $\rho^*= 0.10, 0.14, 0.20,0.26$ and $0.30$ are shown.}
\label{reescala}
\end{center}
\end{figure}

Using this approach we  analyzed the influence of interparticle separation on the decoupling between
translational and non-translational motion. Figure \ref{reescala} shows $T^*_{trans}$ vs $T^*_{ntrans}$ comparing different  
values of $\lambda^*$ for several densities. For small $\lambda^*$, a linear behavior $T^*_{trans} = a^{(\rho)}T^*_{ntrans}$ is
observed but only for a certain interval of densities. As the density increases $a(\rho) \rightarrow 1$. According to 
the definitions described above, this means that, for small interparticle separation, the system can easily 
perform a non-translational motion without performing a translational one. As the  density and the total temperature 
increase the motions of the system become more  correlated. For higher values of $\lambda^*$ the decoupling is not
observed. For all densities $T^*_{trans} = T^*_{ntrans}$. 
In these cases the non-translational motion is strongly dependent (coupled) on translational motion. 

\begin{figure}[h!tb]
\begin{center}
\includegraphics[clip=true,scale=0.4]{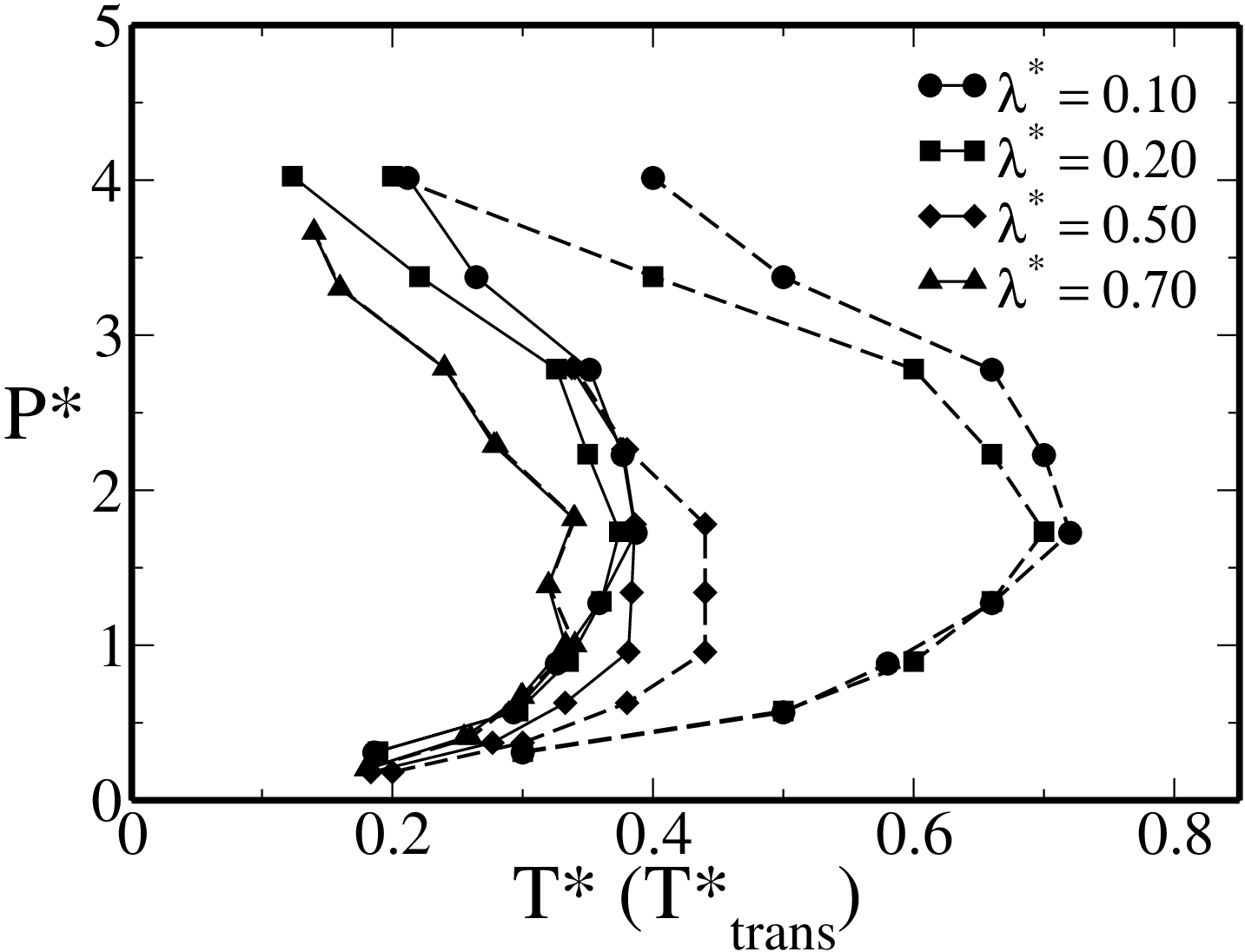}
\caption{Solid-fluid phase boundary for different $\lambda^*$. The solid lines represent the solid-fluid phase boundary rescaled with 
translational temperature, the dashed lines represent the original phase boundary.}
\label{trans}
\end{center}
\end{figure}

For a deeper analysis of the effects  linked to the different degrees of freedom we rescaled the pressure temperature phase diagram
using the translational and non-translational temperature. Figure \ref{trans}(a) shows that the rescaled solid-fluid phase boundary 
with respect the translational temperature moves backwards in such a way that all the curves get closer, almost collapsing. However, when we rescale 
the phase boundary with non-translational function we could observe the opposite behavior. The curves moves forth and they separate 
further from each other.

The behavior of the solid-fluid boundary, as well as the anomalous region  in the pressure-temperature phase diagram are, in fact, 
sensitive to translational temperature. Indeed, in systems where molecules are able to perform non-translational movements without perform 
translational ones, as in the case of small $\lambda^*$,  the dimers interacts with the other dimers in such a way that their center
of mass stay ordered in a solid lattice. Thus, the 
translational temperature acts like the true effective temperature.

Systems with smaller interparticle separation $\lambda^*$ present a larger difference between the translational temperature and the 
formal temperature than systems with larger $\lambda^*$, due to the decoupling of translational and non-translational degrees of freedom.
This difference is responsable for the apparent non-monotonic behavior of the region of solid phase.

\begin{figure}[h!tb]
\begin{center}
\includegraphics[clip=true,scale=0.4]{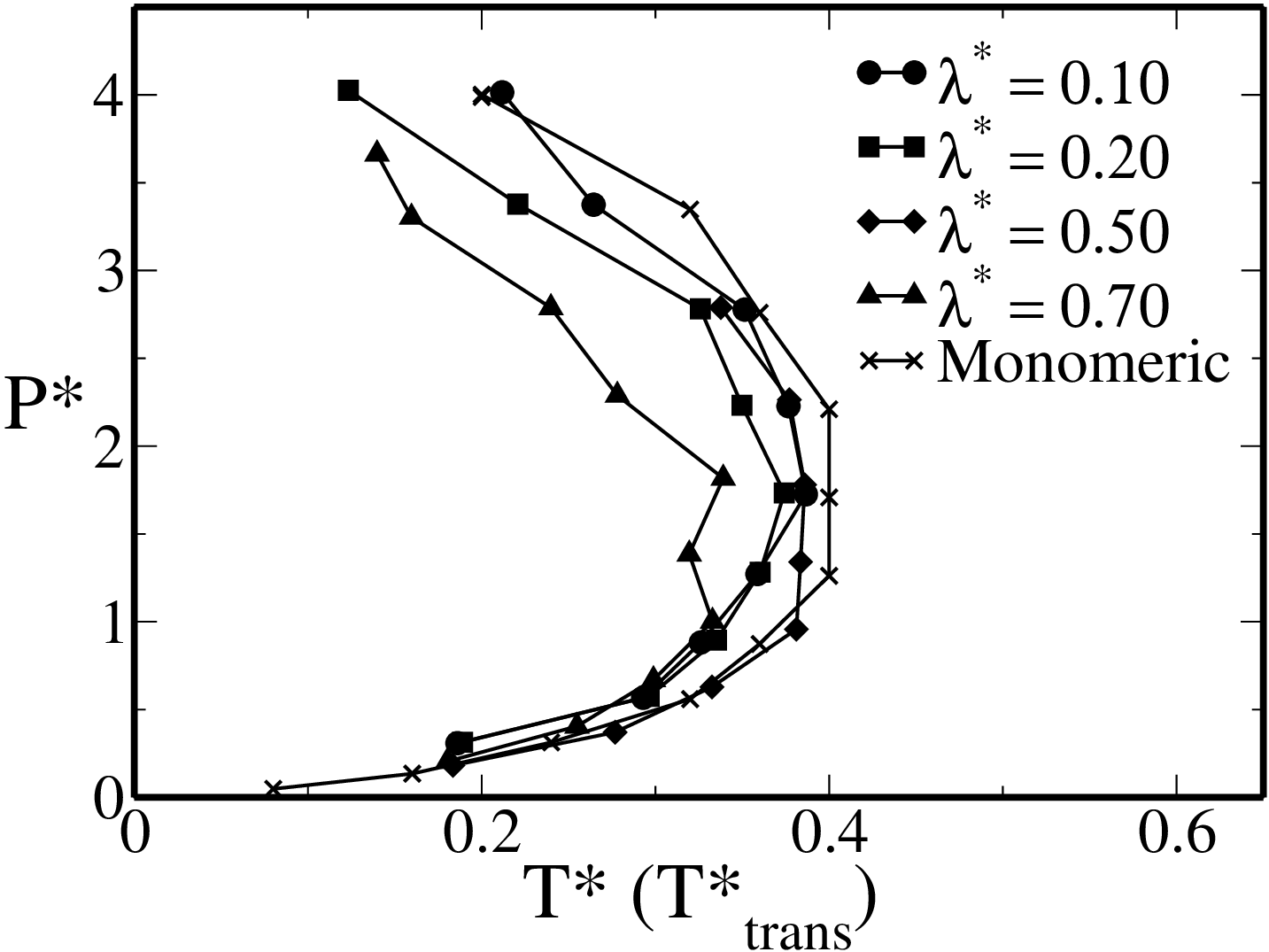}
\caption{Solid-fluid phase boundary for different $\lambda^*$ reescaled with translational temperature and 
Solid-fluid phase boundary for the monomeric system.}
\label{trans2}
\end{center}
\end{figure}

In order to highlight the effects due the translational degrees of freedom, figure \ref{trans2} shows the solid-fluid phase 
boundary rescaled using the translational temperature for several $\lambda^*$ and the solid-fuid phase boundary for the monomeric 
system.

We can observe that, taking in to account only translational movements, the introduction of an anisotropy makes the region of solid
phase lightly move backwards in temperature. We also observe that $\lambda^* = 0.10, 0.20$ and $0.50$ almost collapsed in one region 
indicating the solid-fluid phase boundary depends effectivelly on the translational part of the temperature. Therefore
the unexpected non-monotonic  behavior is due the non-translational motion as we predicted before. Small
dimeric particles can easily  perform a non-translational motion and therefore, despite having a formal high temperature, behave as a solid, due
to their low translational temperature.. 

The phase boundary for $\lambda^* = 0.70$ does not collapsed like the other ones. The reason behind this observation
remains still an open question.

Despite of $\lambda^* = 0.70$ we can affirm that the non-monatomic behavior is essentially  due the difference between translational
temperature and the formal temperature.

\section{Conclusion}

We performed molecular dynamics with rigid dimers in which each monomer interacts with monomers 
from other dimers through a core-softened shoulder potential in order to study the influence of anisotropy
in anomalous regions and solid-fluid phase boundary. We carried out simulations for interparticle
separation $\lambda^* = 0.10, 0.20, 0.50$ and $0.70$ and we observed an unexpected non-monotonic 
behavior where the solid phase shrinks with increase of $\lambda^*$.

Through the analysis of velocity and orientational autocorrelation functions and the mean square displacement
we formulated a hypothesis that decoupling between translational and non-translational motion would be
responsible for  the shrinking of the region of solid phase in the phase diagram.

In order to test these hypothesis we defined the translational and non-translational temperatures 
obteined from the kinetic energy as tools to
evaluate the role of translational and non-translational degrees of freedom. We obtained that for 
simulations with small values of
$\lambda^*$ the non-translational temperature is much higher than the translational one. 
For larger values of $\lambda^*$ the temperatures are
correlated independently of density and temperature. These results are consistent with the non-monotonic
behavior observed.

We rescaled the  pressure temperature phase diagram in order to obtain a pressure vs translational 
temperature phase diagram and observed the behavior of solid phase. 
We argue that translational temperature is the true effective temperature and that the non-monotonic
behavior of solid-fluid phase boundary is due the difference between translational temperature and
the formal temperature. We observed that the introduction of anisotropy leads to a smaller region of solid phase 
when compared with monomeric system and that the solid-fluid phase boundary for $\lambda^* = 0.10, 0.20$ and $0.50$ collapsed on 
one region of phase diagram.

Despite of that we still have open questions since the phase boundary for $\lambda^* = 0.70$ did not 
collapsed like the others. The reason behind that could be related to the size of the dimer. Nevertheless we
succeeded to explain most part of the reasons that led to the non-monotonic behavior of solid-fluid boundary 
which lead us one step forward in the understanding of more complex systems.

%\bibliography{Biblioteca}{}

\end{document}